\documentclass[conference,compsoc]{IEEEtran}
\usepackage{makecell}
\usepackage{longtable}
\usepackage{lscape}
\usepackage{xcolor}
\usepackage{graphicx}
\usepackage{adjustbox}
\usepackage{tabularx}
\usepackage{cite}
\usepackage{booktabs}

\begin{document}

\title{A Review on the Security Vulnerabilities of the IoMT against Malware Attacks and DDoS}

\author{
\IEEEauthorblockN{Lily Dzamesi, Nelly Elsayed}
\IEEEauthorblockA{School of Information Technology,
University of Cincinnati\\
Ohio, United States\\
dzamesly@mail.uc.edu, elsayeny@ucmail.uc.edu}
}

\maketitle
\begin{abstract}

 The Internet of Medical Things (IoMT) has transformed the healthcare industry by connecting medical devices in monitoring treatment outcomes of patients. This increased connectivity has resulted to significant security vulnerabilities in the case of malware and Distributed Denial of Service (DDoS) attacks. This literature review examines the vulnerabilities of IoMT devices, focusing on critical threats and exploring mitigation strategies. We conducted a comprehensive search across leading databases such as ACM Digital Library, IEEE Xplore, and Elsevier to analyze peer-reviewed studies published within the last five years (from 2019 to 2024). The review shows that inadequate encryption protocols, weak authentication methods, and irregular firmware updates are the main causes of risks associated with IoMT devices. We have identified emerging solutions like machine learning algorithms, blockchain technology, and edge computing as promising approaches to enhance IoMT security. This review emphasizes the pressing need to develop lightweight security measures and standardized protocols to protect patient data and ensure the integrity of healthcare services.
 
\end{abstract}


\IEEEpeerreviewmaketitle

\section{Introduction}

\begin{figure}[t]
	\centerline{\includegraphics[width=8cm, height= 6 cm]{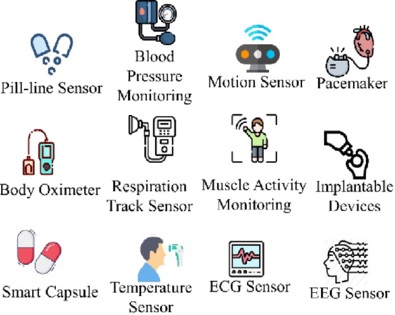}}
	\caption{Examples of IoMT devices.\\
 }
 \label{devices}
\end{figure}

The Internet of Medical Things (IoMT) is a specialized subset of the broader Internet of Things (IoT). It integrates medical device applications to connect with healthcare IT systems through online networks. Devices such as wearable health monitors and advanced imaging systems usually gather and analyze health-related data for continuous monitoring and remote interventions. \cite{lopez2023comprehensive, nair2023internet,bajaj2021healthcare,joyia2017internet,ghubaish2020recent}The remarkable advancements in IoMT have led to improved treatment of patients by allowing healthcare providers to offer more personalized and timely treatments. This includes wearable sensors that track vital signs and implantable devices that monitor critical health conditions. IoMT has become a crucial part of modern medical practices enhancing decision-making capabilities. Figure~\ref{devices} shows an example of IoMT devices~\cite{rani2023federated}.

\begin{figure}[t]
	\centerline{\includegraphics[width=8cm, height= 11.5 cm]{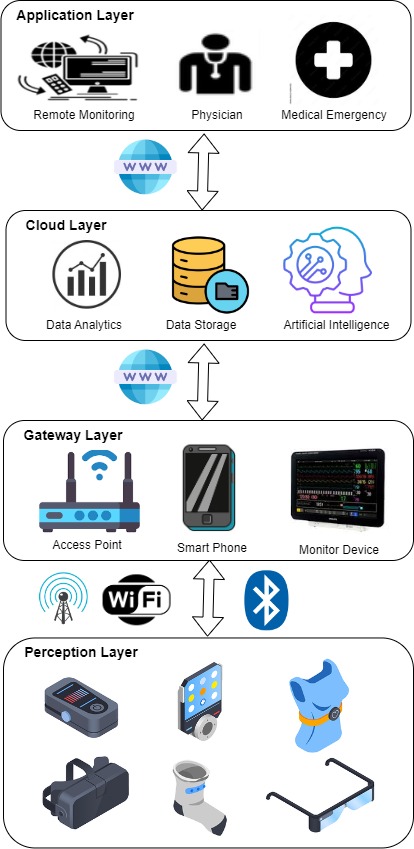}}
	\caption{The IoMT system architecture layers diagram.\\
 }
	\label{IoMT_layers}
\end{figure}

The architecture of IoMT systems comprises four distinct layers: the perception layer, the gateway layer, the cloud layer, and the application layer. The Perception layer in the Internet of Medical Things (IoMT) architecture carries data from medical devices and sensors, including wearables, implanted devices, and traditional medical equipment \cite{hireche2022security,elsayed2023machine,joyia2017internet,hasan2021lightweight}. These devices allow for continuous monitoring of vital signs and communication through protocols such as Bluetooth Low Energy (BLE), Zigbee, and Near Field Communication (NFC)~\cite{tacskin2021container} processing capabilities for preliminary data analysis, like noise reduction and basic analytics, are included in this layer. For example, wearable gadgets can assess heart rate variability and notify the user or healthcare professional of unusual patterns indicating a health risk. The Gateway layer of the IoMT architecture is responsible for aggregating the data that has been collected from the medical devices in the perception layer and serving as a bridge between the cloud and medical devices, ensuring safe data transmission \cite{koutras2020security}. It converts communication protocols such as the Message Queuing Telemetry Transport (MQTT) and Constrained Application Protocol (CoAP) to enable efficient data transmission. The main functions of the gateway layer are the data standardization, data pre-processing, and data transmission~\cite{al2020intelligence}.
the cloud layer in the architecture stores data and analyzes it using cloud computing. It integrates large datasets while utilizing RESTful APIs for seamless communication and scalability in handling growing data volume from IoMT devices \cite{kumar2020secure}. 
The final tier of IoMT architecture is the application layer, where the end-users interact with the system \cite{hatzivasilis2019review}. It supports applications like remote patient monitoring and health management tools to improve healthcare delivery through processed data from the cloud layer. The layer offers intuitive interfaces of dashboards and real-time monitoring alerts for healthcare providers and patients. It uses the HTTP/HTTPS communication protocols for web applications and mobile-specific protocols to provide secure data sharing while adhering to patient privacy and security standards~\cite{dwivedi2022potential}. Finally, the applications in this layer address various healthcare needs while leveraging IoMT's capabilities to improve treatment through secure access and real-time data-driven insights. Figure~\ref{IoMT_layers} shows the IoMT system architecture layers diagram.

Adapting IoMT devices improves the quality of healthcare system monitoring and diagnoses. However, it has increased security vulnerabilities and is a potential target for cyberattacks. According to a survey by Papaioannou et al. (2022) \cite{papaioannou2022survey} indicated most malicious actors exploiting security vulnerabilities use malware infections and Distributed Denial of Service (DDoS) attacks to compromise patient data privacy and the functionality of medical devices.

The vulnerabilities in healthcare systems are often caused by relying on outdated technology that lacks proper encryption and authentication mechanisms Adil et al. (2019) \cite{adil2023covid}. Many IoMT devices were not designed with robust security, which makes them susceptible to cyberattacks. Additionally, Kioskli et al. (2021)~\cite{kioskli2021landscape} emphasize that outdated firmware on these devices exacerbates the risks as it often contains unpatched vulnerabilities that attackers can exploit. Malware infections lead to unauthorized data access that can compromise device functionality and even cause life-threatening device malfunctions.

The DDoS attack on IoMT are performed via different Botnets that targets the Healthcare systems~\cite{saif2023feature,elsayed2023iot}. The DDoS attack can affect the IoMT network by exceeding the network traffic, leading to an interruption in the service distribution and response that can cause a critical delay in accessing the healthcare system resources or even crash or freeze the system~\cite{ul2018ddos}, which affects the system availability of system security based on the Confidentiality, Integrity, Availability (CIA) triad~\cite{azumah2021deep}. For example, during the COVID-19 pandemic, the healthcare systems experienced a tremendous increase in DDoS during the COVID-19 pandemic, which caused a significant network overload, access delay, and denial~\cite{zhou2024statistical}.

The integrity of patient data and the reliability of medical devices are at risk. This paper provides a systematic literature review that addresses these concerns by providing an in-depth analysis of current research trends and mitigation strategies related to the security vulnerabilities of IoMT devices over the last five years. In addition, this paper aims to identify the major challenges facing IoMT device security and investigate potential mitigation strategies.


\section{Method}
\begin{figure}[t]
	\centerline{\includegraphics[width=8.5cm, height= 8 cm]{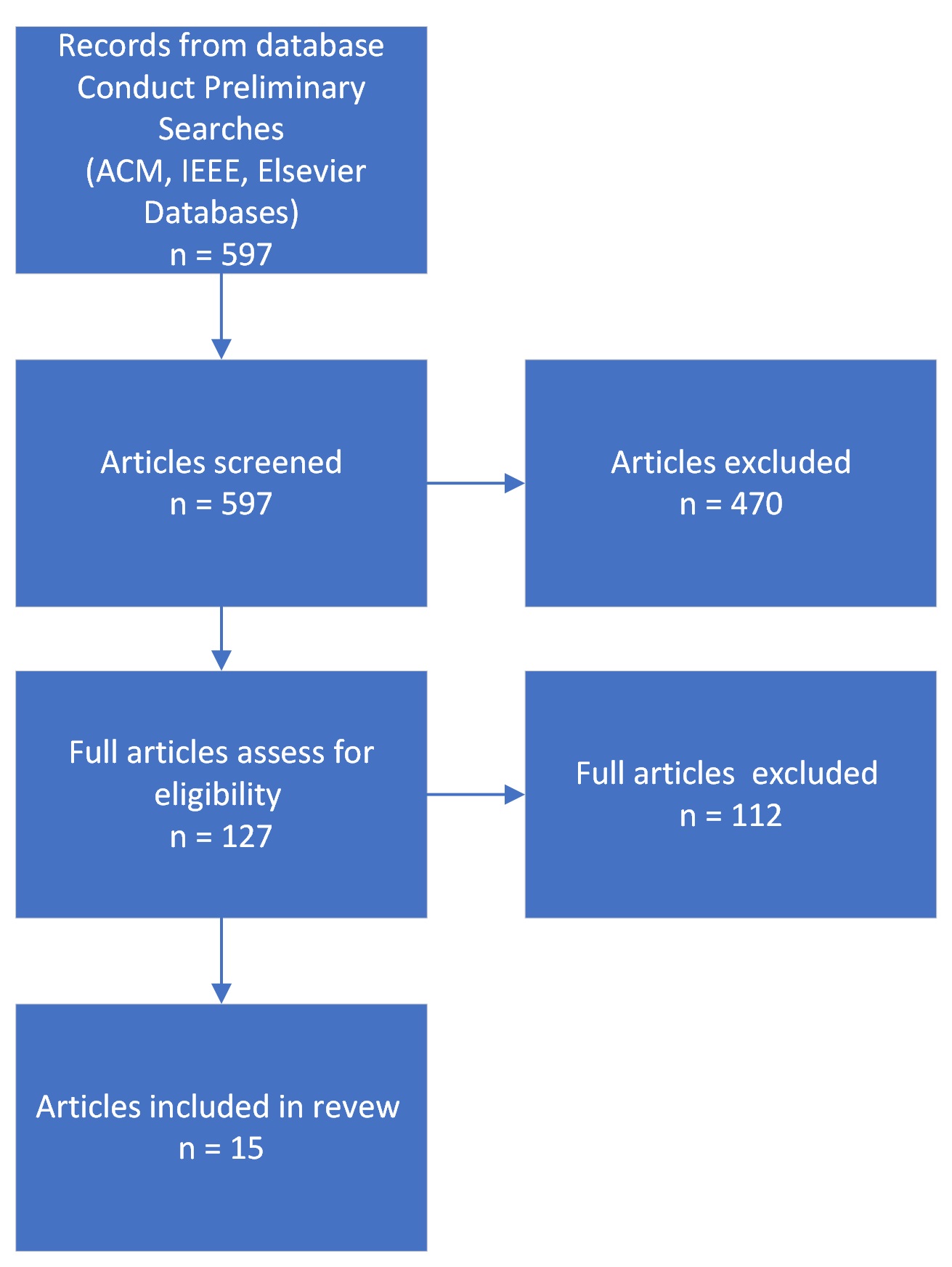}}
	\caption{The workflow used in our systematic search process for IoMT security literature.\\
 }
	\label{systematic}
\end{figure}

In order to address our research questions on the prevailing trends and mitigation strategies for security vulnerabilities in Internet of Medical Things (IoMT) devices concerning malware and Distributed Denial of Service (DDoS) attacks, we conducted a comprehensive systematic literature review as shown in Figure~\ref{systematic}. The following sections detail the methodologies and approaches employed during conducting this literature review.

\subsection{Search Process}
The literature search used three major databases: ACM Digital Library, IEEE Xplore, and Elsevier. These databases were selected to cover many high-quality peer-reviewed articles and conference proceedings. The following search query was used across all databases: 
\begin{itemize}
    \item "query": AllField:(IoMT) AND AllField:(malware /DDoS*) AND (Attacks).

\end{itemize}

\begin{figure}[t]
	\centerline{\includegraphics[width=8.5cm, height= 5.5 cm]{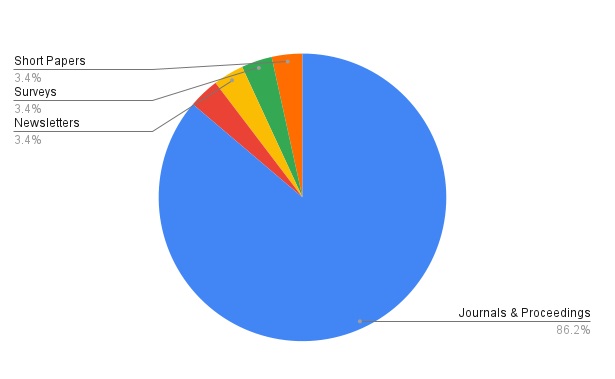}}
	\caption{Distribution of paper sources.\\
    Alt: The diagram shows the distribution of paper sources that have been used at the systematic review.}
	\label{fig:papersSource}
\end{figure}

This search yielded a total of 586 papers: 71 from ACM Digital Library, 27 from IEEE Xplore, and 499 from Elsevier. The papers were classified into various types, including journal articles, conference proceedings, review articles, and book chapters. Figure~\ref{fig:papersSource} shows the distribution of the paper sources.

The search criteria were designed to capture a comprehensive spectrum of literature focusing on security vulnerabilities in IoMT devices, with a specific emphasis on malware and DDoS attacks. By leveraging these diverse databases, we aimed to ensure a robust and inclusive review of pertinent research findings and insights.

\subsection{Inclusion and Exclusion Criteria}
The papers were screened according to established inclusion and exclusion criteria. Studies were required to explicitly address the security vulnerabilities of IoMT with a focus on malware and DDoS attacks. Only peer-reviewed journal articles and conference proceedings published within the last five years were included in this analysis. Studies either addressing IoMT security vulnerability nor published outside the designated date range and not written in English were excluded from the review.

We implemented particular requirements for inclusion and exclusion in order to maintain the relevance and quality of the reviewed studies.

\subsubsection{Inclusion Criteria}
The inclusion criteria that have been adopted in this paper:
\begin{itemize}
    \item Studies specifically addressing the security weaknesses of IoMT devices.
    \item Articles from peer-reviewed journals and proceedings from conferences.
    \item Studies published within the last five years Research published in the past five years (i.e. from 2019 to 2024).
    \item Papers written in English.
\end{itemize}

\subsubsection{Exclusion Criteria}
\begin{itemize}
    \item Research that did not specifically examine security concerns in the Internet of Medical Things (IoMT) or did not concentrate on malware or Distributed Denial of Service (DDoS) attacks.
  \item  Papers that fall outside the designated date range or are not published within a peer-reviewed context. 
\item Papers in languages other than English.
\end{itemize}

\subsection{Quality Assessment}
Each selected study was evaluated to determine its methodological contribution to the field of IoMT security. This assessment ensured that only exceptional papers presenting significant evidence and insights regarding malware and DDoS attacks on the Internet of Medical Things (IoMT) were included. The quality assessment helped to exclude papers with methodological deficiencies and those focused on unrelated topics.

\subsection{Data Collection}
We collected data from selected studies by categorizing them into four key areas based on the publications that met our inclusion criteria. The resulting database served as the foundation for our analysis. Our data collection process involved extracting relevant information from each publication to ensure a clear understanding of their content and contributions to the field, as demonstrated in Figure 2. Below, we outline the key aspects of this process:

\begin{itemize}
    \item \textit{Type of Publication:} We classified the publications as either journal articles or conference proceedings. This helped us identify the contexts in which these studies were presented and the credibility associated with each publication type.
\end{itemize}
\begin{itemize}
    \item \textit{Publication Venue:} The journals and conferences where the studies were published were recorded to assess their impact and relevance within the cybersecurity community. We focused on reputable venues such as IEEE, ACM, and Elsevier journals.
\end{itemize}
\begin{itemize}
    \item \textit{Year of Publication:} By recording the year of publication, we could track emerging trends over time, particularly regarding the evolution of malware and DDoS attack methodologies and the advancement of security measures.
\end{itemize}
\begin{itemize}
    \item \textit{Technology Utilized:} Each study was analyzed for the technological frameworks and platforms they focused on, such as encryption methods, machine learning algorithms, blockchain technology, and edge computing. This helped us map the technological landscape within IoMT security.
\end{itemize}
\begin{itemize}
    \item \textit{Security Vulnerabilities:} We documented the specific security vulnerabilities identified in each study, focusing on malware infections and DDoS attacks. This categorization allowed us to synthesize the critical risks faced by IoMT devices.
\end{itemize}
This systematic process ensured a thorough literature review and enabled a comprehensive understanding of IoMT security vulnerabilities.

\subsection{Data Extraction Process}
Once the papers were categorized, we extracted the key information from each study, which included the following:
\begin{itemize}
    \item \textit{Research Focus:} The primary security vulnerabilities or mitigation strategies addressed.
    \item \textit{Methodological Approach:} The research methods used (e.g., case studies, simulations, empirical data).
    \item \textit{Key Findings:} Specific insights or proposed solutions from the study.
    \item \textit{Limitations:} Any noted limitations or challenges in the proposed security strategies.
\end{itemize}

\subsection{Synthesis of Findings}
We extracted data and synthesized it to identify recurring themes and challenges across the studies. We focused on the susceptibility of IoMT devices to malware and DDoS attacks, the effectiveness of various mitigation strategies, and the gaps in current security research. The systematic organization of data enabled us to draw meaningful conclusions about the state of IoMT security and its future directions.

\section{Synthesis of Findings}
The systematic literature review found several common themes and important findings about the security vulnerabilities of Internet of Medical Things (IoMT) devices. The studies highlighted the widespread occurrence of malware infections and Distributed Denial of Service (DDoS) attacks as the most significant security threats. The review also provided insights into the most effective methods used to address these security vulnerabilities, as indicated in Table 1. The results are categorized into the following key areas:


\begin{table*}[htbp]
\centering
\caption{Mitigation Approaches and Vulnerabilities in IoMT Systems.}
\begin{tabularx}{\textwidth}{|X|X|X|X|X|X|}
\hline
\textbf{Paper} & \textbf{Mitigation Approach} & \textbf{Description} & \textbf{Advantages} & \textbf{Limitations} & \textbf{Devices Used } \\ \hline
Adil et al.~\cite{adil2023covid} (2023) & Cryptography & Encryption techniques to secure IoMT data, particularly in pandemic healthcare settings. & Protects data confidentiality and integrity. & Increased processing overhead requires efficient key management. & IoMT devices in COVID-19 monitoring systems. \\ \hline
Saxena and Mittal~\cite{saxena2022internet} (2022) & Cryptography & Secure communication channels using encryption and authentication for IoMT data transfers. & Secures both at-rest and in-transit data. & Requires high computational resources for IoMT devices with limited power. & Wearable health monitoring devices. \\ \hline
Wazid and Gope~\cite{wazid2023backm} (2023) & Blockchain & A blockchain-based system for secure data sharing and storage in IoMT applications. & Decentralized, tamper-proof, and offers better access control. & High computational overhead; scalability issues in large IoMT ecosystems. & E-health applications with interconnected medical devices. \\ \hline
Brass and Mkwashi~\cite{brass2022risk} (2022) & Risk Assessment & Framework for identifying and classifying vulnerabilities in medical IoMT software. & Helps prioritize vulnerabilities and assess risks efficiently. & Limited to specific device types and software categories. & Software in connected medical devices. \\ \hline
Kioskli et al.~\cite{kioskli2021landscape} (2021) & Machine Learning & Utilizes machine learning to detect patterns indicating potential DDoS attacks on IoMT networks. & Real-time detection, adaptable to evolving attack methods. & Requires large datasets for training; false positives possible. & Smart healthcare systems, IoMT devices in hospitals. \\ \hline
Bhutia et al.~\cite{bhutia2022ddos} (2022) & Machine Learning & Machine learning algorithms to identify and block DDoS attacks in IoMT environments. & Adaptable and highly accurate with large datasets. & Computationally expensive, resource-intensive for low-power devices. & Medical diagnostic devices connected to IoT. \\ \hline
De Michele and M. Furini~\cite{de2019iot} (2019) & Cryptography & Lightweight cryptographic methods for IoT and IoMT healthcare networks. & Minimal overhead on processing; secure communications. & Limited encryption strength, vulnerable to advanced attacks. & Low-power IoMT devices, including portable monitors. \\ \hline
Hameed et al.~\cite{hameed2021systematic} (2021) & Machine Learning & A review of machine learning techniques to detect malware and unauthorized access in IoMT systems. & Effective for detecting zero-day attacks. & High false positive rates; requires significant training data. & IoMT devices integrated into hospitals. \\ \hline
Da Costa et al.~\cite{da2019internet} (2019) & Machine Learning & Intrusion detection systems (IDS) for IoMT using machine learning-based analysis. & Fast response to threats, proactive security measures. & Expensive to implement and maintain on resource-constrained IoMT devices. & Remote patient monitoring systems. \\ \hline
Chatterjee et al.~\cite{chatterjee2023approach} (2023) & Blockchain & Blockchain solutions for managing sensitive medical data in IoMT networks. & Provides secure, decentralized data management and integrity. & High latency in data processing; issues with scalability for large datasets. & IoMT-enabled e-healthcare systems. \\ \hline
\end{tabularx}
\label{table:mitigation}
\end{table*}

\subsection{Vulnerabilities in IoMT Devices}
The review revealed that many IoMT devices suffer from weak security configurations, making them particularly vulnerable to cyberattacks \cite{brass2022risk}. If not addressed, these vulnerabilities can compromise patient safety and healthcare services' integrity. 

\begin{itemize}
    \item \textit{Inadequate Encryption:} Many studies have identified inadequate encryption protocols as a significant vulnerability in IoMT devices, potentially resulting in unauthorized access to sensitive patient information. Outdated encryption methods increase the vulnerability of devices to interception and manipulation by attackers \cite{adil2023covid}. Weak encryption in pandemic healthcare monitoring systems has resulted in vulnerabilities in the protection of patient information.

    \item \textit{Weak Authentication Mechanisms:} IoMT devices utilize fundamental or legacy authentication methods which are very susceptible to attacks. Weak mechanisms can be exploited by attackers to gain unauthorized access further compromising both the device and the data it manages \cite{papaioannou2022survey}. This is especially alarming in devices that are widely distributed and do not receive adequate security updates.
    
    \item \textit{Limited Processing Power and Irregular Firmware Updates:} The resource-constrained characteristics of several IoMT devices frequently result in the neglect of regular firmware updates, thereby exposing them to newly identified vulnerabilities. The limited processing power also constrains the adoption of more advanced security measures rendering these devices appealing targets for attackers \cite{wazid2023backm}. Numerous IoMT ecosystems exhibit vulnerabilities resulting from delayed updates and insufficient hardware capabilities.
\end{itemize}

\subsection{Malware Attacks}

Malware poses a significant threat to IoMT devices due to delayed updates and inadequate hardware capabilities~\cite{hameed2021systematic}. Malware remains one of the most significant threats to IoMT devices. In many cases, attackers exploit vulnerabilities in outdated or unpatched devices by installing malware that can exfiltrate sensitive healthcare data and services; one such example is the infamous Mirai malware, which transformed IoMT devices into bots to launch large-scale DDoS attacks. According to recent studies~\cite{elhoseny2021security}, over 70\% of IoMT devices currently in use are vulnerable to known malware attacks in exposing healthcare providers and patients to significant risks.

To address this issue, machine learning techniques have shown promise in identifying abnormal behavior patterns that indicate malware infection. This proactive approach enables healthcare providers to respond to threats before they cause irreparable harm. Network devices are overwhelmed with traffic as well as disrupting healthcare operations and preventing timely access to patient data~\cite{saxena2022internet}.

\subsection{Distributed Denial of Service (DDoS) Attacks}
Numerous studies have shown that DDoS attacks can severely impact IoMT networks. These attacks overwhelm devices with traffic, disrupt healthcare operations, and hinder timely access to patient data. This poses a significant risk in emergency medical services where connectivity and response times are crucial~\cite{bhutia2022ddos,da2019internet}. This is especially dangerous in healthcare settings, where timely access to data and services can be a matter of life or death. A well-known case occurred in 2017 when a DDoS attack targeted a major hospital network, halting communication between doctors and life-saving equipment, such as ventilators and heart monitors \cite{al2023cybersecurity}.
The resource-constrained nature of many IoMT devices makes them easy targets for DDoS attacks, as they often lack the computational power to fend off large-scale attacks. Machine learning algorithms, such as classification models, can be used to differentiate between legitimate and malicious traffic in real-time, enabling the system to filter out DDoS attack attempts without disrupting the normal flow of data.

\subsection{Mitigation Strategies}
The review also identified various promising mitigation strategies summarized as:

\begin{enumerate}
    \item{\textit{Machine Learning Algorithms:}} These were highlighted for their ability to detect anomalies in network traffic and identify potential malware and DDoS threats in real-time \cite{saxena2022internet,da2019internet}.

    \item {\textit{Blockchain Technology:}} Blockchain’s decentralized and tamper-proof architecture was found to offer enhanced access control and secure data storage\cite{wazid2023backm,chatterjee2023approach}.

    \item {\textit{Edge Computing:}} Studies indicated that edge computing could help localize threat detection and mitigation in reducing the latency associated with cloud-based security measures\cite{gudala2019leveraging}.
\end{enumerate}

\begin{figure}[t]
	\centerline{\includegraphics[width=8.5cm, height= 5cm]{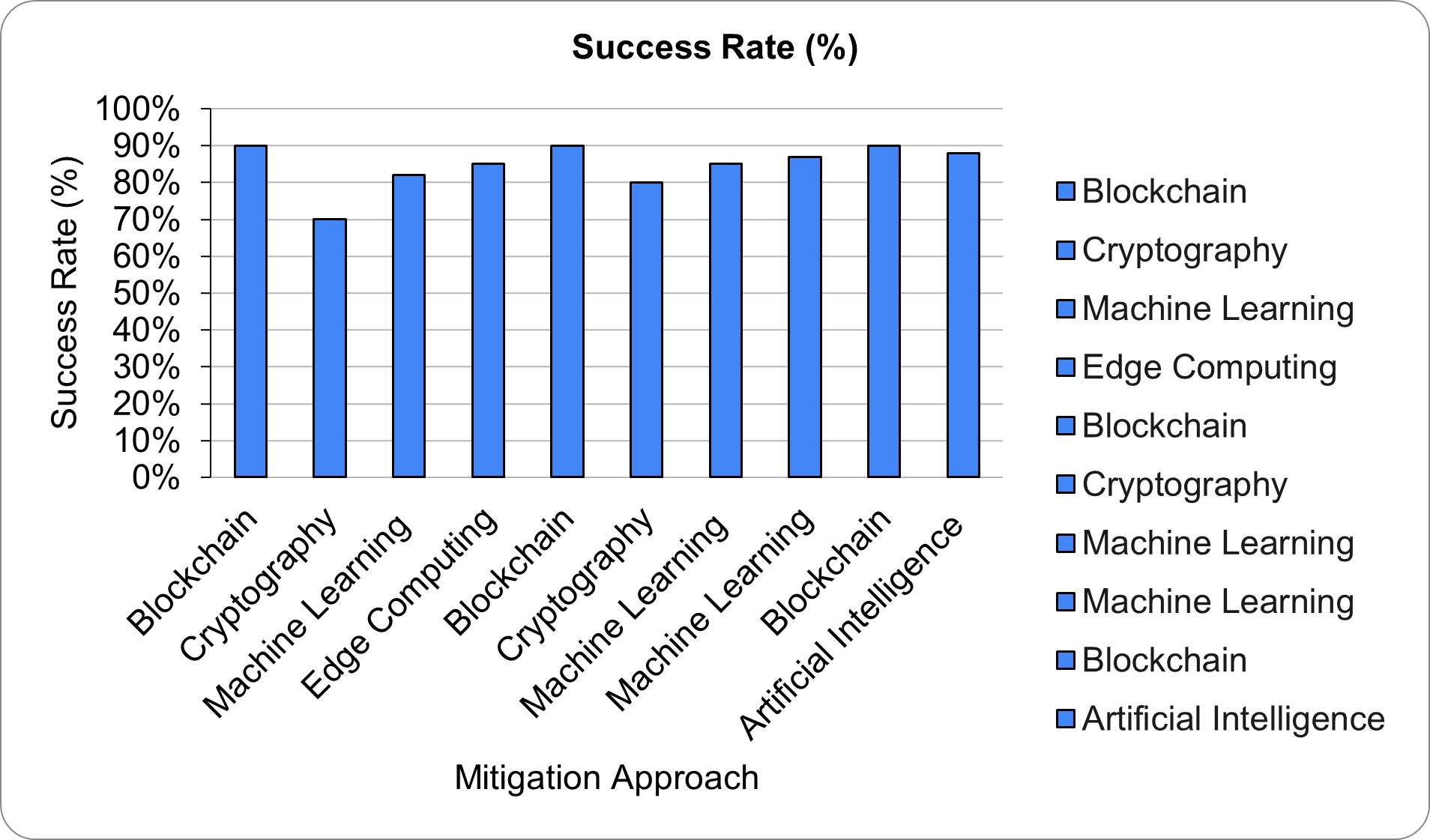}}
	\caption{The success rate.}
	\label{figure5}
\end{figure}

\begin{figure}[t]
	\centerline{\includegraphics[width=8.5cm, height= 5cm]{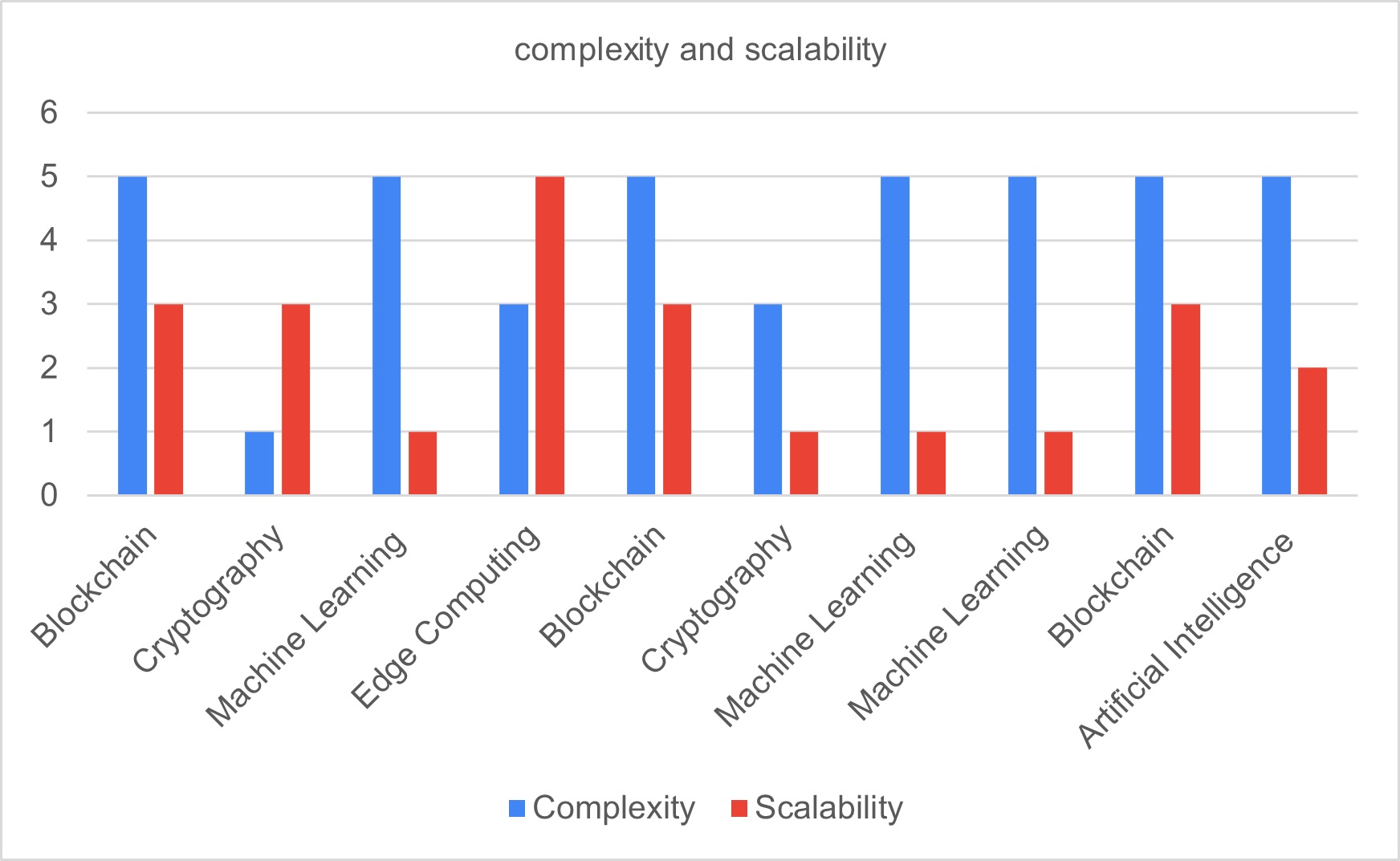}}
	\caption{The complexity and scalability.}
	\label{figure6}
\end{figure}

The analysis of our research papers also provides insights into the effectiveness of different mitigation approaches used to address security vulnerabilities in the Internet of Medical Things (IoMT). These mitigation approaches were evaluated based on three key factors: Success Rate, Complexity, and Scalability. as shown in Figures~\ref{figure5} and~\ref{figure6} .
\begin{itemize}
    \item \textit{Success Rate:}
The success rate of the mitigation approaches ranged from 70\% to 90\%. The highest success rate was achieved by blockchain-based solutions (90\%) as demonstrated by \cite{papaioannou2022survey}, \cite{wazid2023backm}, and \cite{chatterjee2023approach}. These solutions provided decentralized and tamper-proof mechanisms for securing IoMT devices.
Machine learning approaches also showed high success rates, particularly in detecting DDoS attacks and malware, with success rates ranging between 82\% and 88\% (\cite{bhutia2022ddos,da2019internet}).
Cryptographic methods had moderate success rates, with a range of 70\% to 85\% (\cite{adil2023covid,de2019iot,saxena2022internet}).

    \item \textit{Complexity:}
Mitigation strategies varied significantly in terms of complexity. Machine learning and blockchain approaches were generally associated with high complexity due to their computational requirements and the need for large datasets (\cite{papaioannou2022survey}, \cite{gudala2019leveraging}, \cite{bhutia2022ddos}, \cite{da2019internet}).
Edge computing and lightweight cryptography were identified as moderate to low complexity solutions mostly suitable for resource-constrained IoMT devices (\cite{de2019iot}, \cite{brass2022risk}).

    \item \textit{Scalability:}
Edge computing and blockchain-based solutions showed higher scalability, as they can be implemented across diverse IoMT ecosystems and can handle large-scale data without compromising performance (\cite{papaioannou2022survey,brass2022risk,wazid2023backm}).
Machine learning approaches were effective. However, they exhibited lower scalability due to their reliance on high computational power and the need for substantial training datasets (\cite{hameed2021systematic,bhutia2022ddos}).

\end{itemize}


\subsection{Challenges in Implementing Security Measures} 
Despite the promising strategies, many studies noted challenges in implementing robust security protocols:
\begin{itemize}
    \item \textit{High Computational Overhead:} Cryptographic techniques like blockchain were found to require significant processing power, which is difficult to achieve with resource-constrained IoMT devices \cite{elhoseny2021security,rahman2021security}.
    \item \textit{Scalability Issues:} Some solutions, such as machine learning models, seem to require large datasets for training and often struggle with scalability when applied to larger IoMT ecosystems \cite{hameed2021systematic,rahman2021security}.
    
\end{itemize}

\section{Discussion}

This systematic literature review reveals that while IoMT devices offer transformative benefits to the healthcare sector, their growing security vulnerabilities pose a significant concern. As Table 1 highlights, various mitigation strategies have been explored across this literature with differing levels of complexity and scalability, as demonstrated in Figures~\ref{figure5} and~\ref{figure5} . These vulnerabilities present substantial risks to patient safety and the healthcare sector \cite{hameed2021systematic}. The weak encryption protocols and irregular firmware updates found in many IoMT devices make them attractive targets for cybercriminals, threatening the integrity and confidentiality of sensitive patient data.

Machine learning algorithms have emerged as a promising tool for identifying and mitigating threats in real-time IoMT environments \cite{bhutia2022ddos}. Despite their effectiveness in detecting anomalies and malicious behavior, these models often require considerable computational resources and large datasets, which makes them less suitable for resource-constrained IoMT devices. As a result, scalability remains a critical challenge in applying machine learning across broader IoMT ecosystems.

Similarly, blockchain technology offers a decentralized and tamper-proof framework for securing IoMT data \cite{chatterjee2023approach}. However, while blockchain enhances data integrity and access control, the high computational overhead limits its applicability, especially in large-scale healthcare systems with prevalent resource constraints. These limitations underscore the need for more efficient blockchain implementations or alternative lightweight cryptographic techniques that can better align with the hardware limitations of IoMT devices.

Furthermore, the review highlights the importance of developing lightweight cryptographic solutions. Current security methods often overwhelm the limited processing power of IoMT devices, becoming impractical for widespread adoption \cite{de2019iot}. Future research should prioritize designing security measures that balance robustness with resource efficiency to ensure broader application in healthcare environments.

The urgency for standardized security protocols emerged as the recurring theme in the study. The lack of frameworks for securing IoMT devices exacerbates interoperability and cross-device communication issues. It is, therefore, prudent to develop global industry standards to mitigate security vulnerabilities and streamline the integration of security measures across IOMT's diverse ecosystems, ultimately improving the reliability and effectiveness of healthcare services.

\section{Conclusion}
The rapid proliferation of Internet of Medical Things (IoMT) devices has significantly boosted healthcare delivery and introduced substantial security issues. Our review identified malware and DDoS attacks as the major threats, often due to weak encryption, outdated authentication methods, and irregular firmware updates.

In this systematic review, several promising strategies to mitigate these risks have been identified. These include machine learning algorithms, which can detect and respond to security threats in real-time blockchain technology to secure and transparently manage data and edge computing which can reduce latency and improve data processing efficiency. All of these could enhance IoMT security.

The implementation of these mitigation strategies is challenged by issues such as computational overhead, scalability problems in managing a large number of devices, and the lack of industry-standard protocols for interoperability. To effectively address these challenges, future research should focus on developing lightweight security measures that are tailored to the constraints of IoMT devices and establishing standardized protocols to improve the interoperability and resilience of healthcare systems.

To ensure a secure future for IoMT in healthcare more collaboration between researchers and device manufacturers is essential. This will mitigate security risks and foster the trust necessary for adopting IoMT devices in critical healthcare environments.



%
\bibliographystyle{ieeetr}
\bibliography{bio.bib}

\end{document}